\documentclass[%
 reprint,
 amsmath,amssymb
 aps,
]{revtex4-1}

\usepackage{graphicx}
\usepackage{dcolumn}
\usepackage{bm}
\usepackage{color}
\usepackage{wasysym}
\usepackage{epstopdf}
\usepackage{verbatim}
\usepackage{natbib}
\usepackage{epsfig}
\usepackage{subfigure}
\usepackage{amsmath}
\usepackage{float}

\begin{document}

\preprint{APS/123-QED}

\title{Long Light Storage Time in an Optical Fiber}

\author{Wui Seng Leong, Mingjie Xin, Chang Huang, Zilong Chen}
\author{Shau-Yu Lan}%
 \email{sylan@ntu.edu.sg}
\affiliation{%
Division of Physics and Applied Physics, School of Physical and Mathematical Sciences, Nanyang Technological University, Singapore 637371, Singapore
}%




\date{\today}

\begin{abstract}
Light storage in an optical fiber is an attractive component in quantum optical delay line technologies. Although silica-core optical fibers are excellent in transmitting broadband optical signals, it is challenging to tailor their dispersive property to slow down a light pulse or store it in the silica-core for a long delay time. Coupling a dispersive and coherent medium with an optical fiber is promising in supporting long optical delay. Here, we load cold Rb atomic vapor into an optical trap inside a hollow-core photonic crystal fiber, and store the phase of the light in a long-lived spin-wave formed by atoms and retrieve it after a fully controllable delay time using electromagnetically-induced-transparency (EIT). We achieve over 50 ms of storage time and the result is equivalent to 8.7$\times$10$^{-5}$ dB $\mu$s$^{-1}$ of propagation loss in an optical fiber. Our demonstration could be used for buffering and regulating classical and quantum information flow between remote networks.

\end{abstract}

\pacs{Valid PACS appear here}
\maketitle

Optical delay lines or optical buffers play important roles in long-distance quantum communication networks for storing, delaying, and, thus, exchanging information between different quantum nodes. The superb performance of optical fibers as transmission lines has driven the consideration of integrating these functionalities into the optical fibers themselves \cite{Th,Geh}. One of the challenges is maintaining the coherence of the medium in the fiber in which the information of light is encoded. The direct use of solid-core materials of telecommunication band fibers via stimulated Brillouin scattering and doped erbium ions to store light has been demonstrated to tens of nanoseconds \cite{Zhu,Sag,Jin,Sag2}. Despite their tens of GHz bandwidth and telecom wavelength operation, the short-lived acoustic waves in the materials and spin coherence of doped erbium in the respective settings limit the performance of the light storage.

Alternatively, interfacing long-lived atomic spin states with optical fibers could provide a route to achieving long storage time. Loading room temperature atomic vapor into a hollow-core photonic crystal fiber \cite{Spr} and cold atoms near the surface of an optical nanofiber \cite{Gou} have achieved storage times of about tens of nanoseconds and few microseconds, respectively. They are limited by the transit time of atoms through the optical modes. Confining atoms in the evanescent field of an optical nanofiber \cite{Say,Cor} and guided field of a hollow-core fiber \cite{Bla,Pet} could eliminate decoherence from transit time but introduce additional decoherence from the confining fields. Both systems have shown storage time of about a few microseconds. To overcome the decoherence, we use a state insensitive optical potential to confine $^{85}$Rb atoms in the guiding mode of a 4-cm-long hollow-core photonic crystal fiber \cite{Deb,Alh}. The phase of the optical pulse is mapped onto the atomic spin wave formed by a pair of long-lived hyperfine ground states and retrieved with a controllable delay using EIT. We demonstrate light storage over 50 ms with a bandwidth of 1 MHz.

\begin{figure}[H]
\includegraphics[scale=0.33]{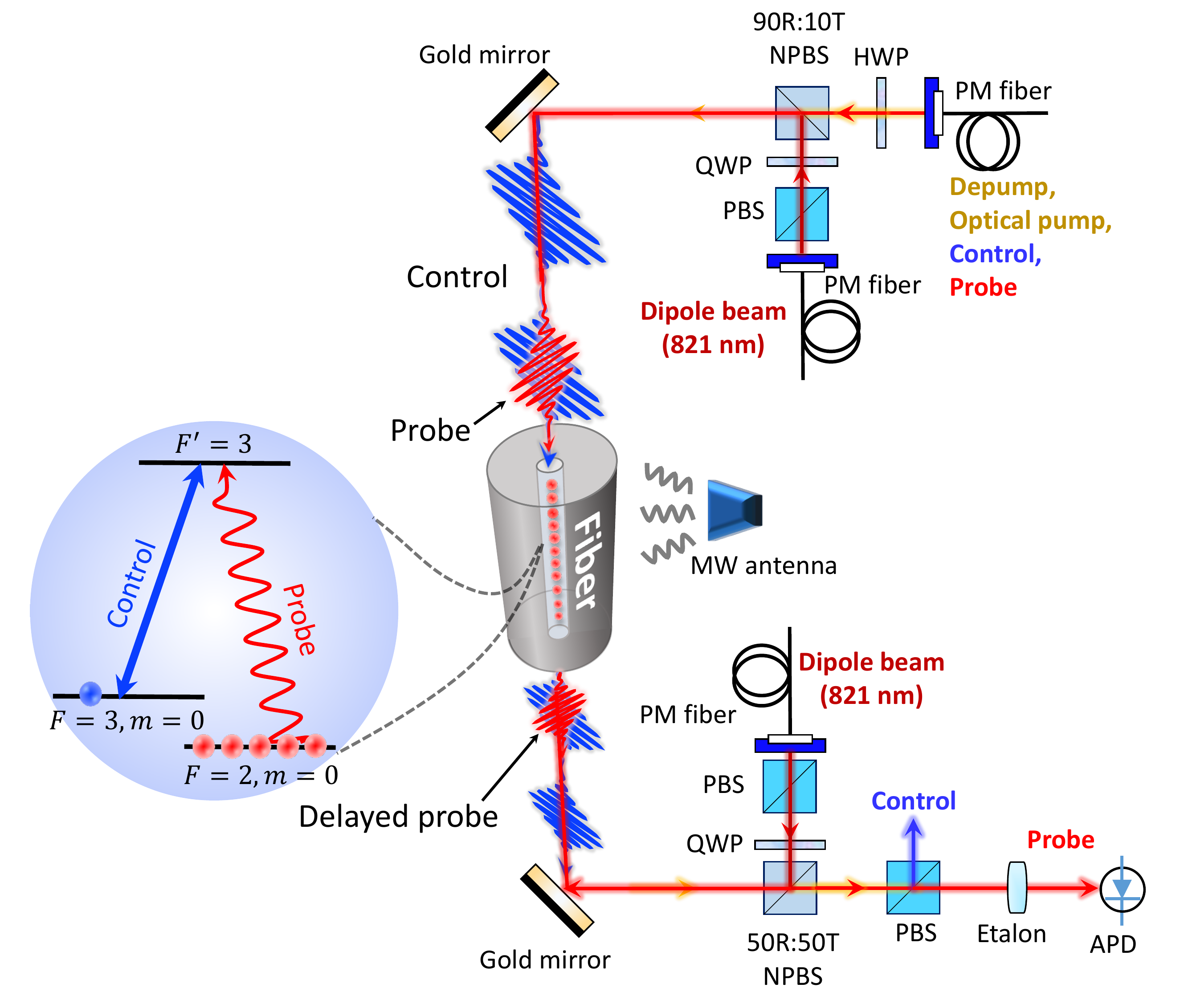}
\caption{\label{fig:epsart} Details of the experimental setup. NPBS: non-polarizing beam splitter. PBS: polarizing beam splitter. HWP: half-wave plate. QWP: quarter-wave plate. PM fiber: polarization-maintaining fiber. APD: avalanche photodiode. The QWPs are used to adjust the ellipticity of the dipole beams to induce the vector light shift. In order to combine the linearly polarized EIT beams and the elliptically polarized dipole beam, we use a 90$\%$ reflection and 10$\%$ transmission NPBS. A 50:50 NPBS is then used to separate the EIT beams from the counter-propagating dipole beam. We use a temperature-stabilized solid-state etalon to minimize leakage of the control beam to the APD. The depump light on the $F$=3 to $F$'=3 D2 line is used to prepare the atoms in the $F$=2 state. The optical pump beam is resonant on the $F$=2 to $F$'=2 D1 line. The first control pulse is for converting the probe pulse into an atomic spin-wave and the second is for retrieving it.}
\end{figure}

\begin{figure*}
\includegraphics[scale=0.8]{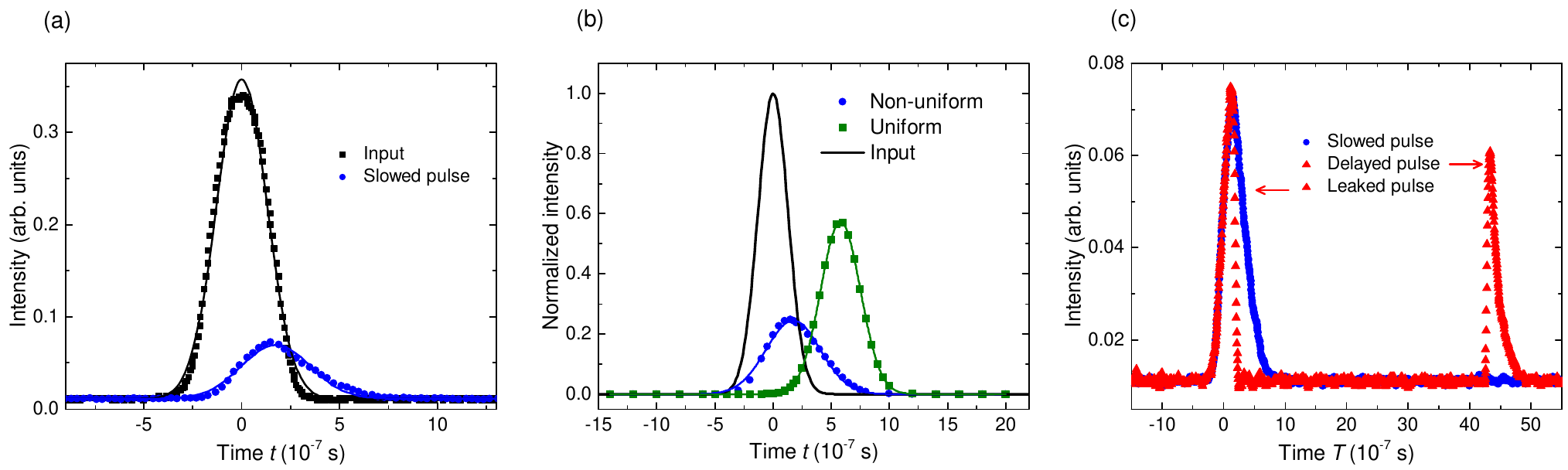}
\caption{\label{fig:epsart} Input, slowed, and delayed light pulses through the hollow-core fiber. (a) Measured slowed light when the control beam is permanently on. The input pulse (black squares) is fitted by a Gaussian function with a FWHM of 304 ns, and the slowed pulse (blue circles) is fitted by a Gaussian function with a FWHM of 415 ns. The peak of the fitted Gaussian function of slowed light is delayed by 173 ns. (b) Numerical calculations of slowed pulses. The blue circles are calculations of the slowed pulse with non-uniform profile of the control beam, the probe beam, and the atom density in the fiber. The green squares are the calculations of the slowed light with uniform EIT beam profiles and atom density distribution. (c) The measured temporal profiles of the leaked and delayed light pulses. The control pulse is switched off when the probe pulse is propagating inside the atomic medium. The bandwidth of the delayed pulse is defined from its FWHM of 160 ns. }
\end{figure*}

A cold $^{85}$Rb atomic ensemble is prepared by a magneto-optical trap (MOT) $\sim$5 millimeters above a 4-cm-long open-end hollow-core photonic crystal fiber, as shown in Fig. 1. After sub-Doppler cooling, the atomic ensemble is loaded into an optical trap guided by the fiber with a temperature of 10 $\mu$K. The wavelength of the optical dipole beam is 821 nm with 250 mW power, and the measured optical depth ($D$) is about 40 by probing all the Zeeman states of the $^{85}$Rb D2 line $F$=3 to $F$'=4 transition. When the atoms are in the fiber, we optically pump them into the $F$=2, $m$=0 state by a $\pi$-polarized light and a repump light coupled through the fiber. The quantization axis is defined by a 2 G magnetic field along the fiber axis and the magnetic field is also used to lift the degeneracy of the Zeeman sublevels.

Our EIT scheme is formed by a three-level $\Lambda$ configuration \cite{Fle} operating on the hyperfine clock states of $^{85}$Rb for long coherence times. The probe beam is resonant on $|$1$\rangle$=$|F$=2, $m$=0$\rangle$ to $|$3$\rangle$=$|F$'=3, $m$=$\pm$1$\rangle$ and the control beam is resonant on $|$2$\rangle$=$|F$=3 $m$=0$\rangle$ to $|$3$\rangle$. A 795 nm diode laser is tuned to the control beam frequency, and a small portion of the power is split to set up the probe beam by an electro-optical modulator around 3 GHz followed by an etalon to filter out the undesired sidebands. Both beams are perpendicularly linearly polarized, and co-propagate inside the fiber. After exiting the fiber, the probe beam is separated from the control beam by polarization and etalon filtering and detected with an avalanche photodiode. The control and probe beams power inside the fiber are 300 nW and 40 nW, respectively, and the frequencies and the power of both beams are controlled by two independent and phase coherent acousto-optical modulators (AOMs). The 80 MHz radio frequency for the probe AOM is modulated by a Gaussian pulse from a waveform synthesizer. The 80 MHz radio frequency for the control AOM is modulated by a square pulse.

When the atoms are under free fall inside the fiber, we measure the group delay of the probe pulse, as shown in Fig. 2(a). The intensity of the input probe pulse can be approximated with a Gaussian temporal profile of $I_{\textrm{p}}$($t$)=$I_{\textrm{p0}}$exp(-4(ln2)$t^{2}$/$T_{\textrm{p}}^{2}$), where $I_{\textrm{p0}}$ is the maximum intensity, $T_{\textrm{p}}$=304 ns is the FWHM duration, and $t$ is the temporal coordinate. A Gaussian function is fitted to the transmitted probe pulse to determine the group delay $T_{\textrm{d}}$=174 ns with 24$\%$ efficiency of transmission.

\begin{figure*}
\includegraphics[scale=0.8]{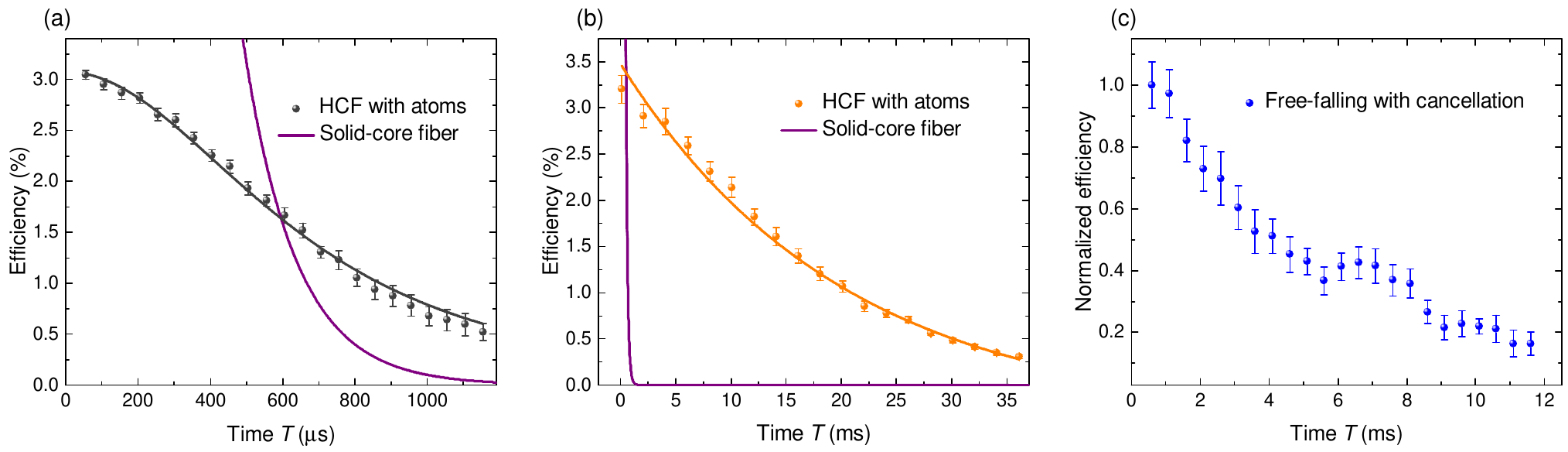}
\caption{\label{fig:epsart} Storage efficiency as a function of time. (a) The grey circles are the measurements of the storage efficiency without light shift cancellation. The function $A$(1+($T$/$T_{\textrm{c}}$)$^{2}$)$^{-3/2}$ is fitted to the data with $T_{\textrm{c}}$=823(10) $\mu$s, where $A$ is the efficiency at $T$=0. The decay time is limited by the differential ac Stark shift of the dipole potential on the atomic spin coherence. For comparison, the curve in purple represents the 0.03 dB $\mu$s$^{-1}$=0.15 dB km$^{-1}$ loss of 1550 nm light propagating in a solid core fiber. The error bars are the standard errors of 40 experimental runs. (b) Measurements of the efficiency with light shift cancellation (orange circles). An exponential decay function is fitted to the data with a 1/$e$ time of 20(1) ms. The curve in purple is the same as in (a). The error bars are the standard errors of 40 experimental runs. (c) Measurements of normalized storage efficiency for a free-falling atomic spin-wave with light-shift cancellation. The control and probe pulses are sent when atoms are in motion inside the fiber.}
\end{figure*}

We numerically calculate the group delay and the efficiency of the transmitted light by integrating over the inhomogeneous profiles of the atomic cloud in the dipole trap, the probe beam, and the control beam inside the fiber from the center of the fiber to the wall of the fiber $W$=31.5 $\mu$m \cite{Hsi}. The intensity of the probe pulse after transmitting through the atomic ensemble is obtained as

\begin{equation}
\begin{split}
\int_{0}^{W} 2\pi r\times(\frac{\Omega_{\textrm{p}}}{\beta}\times\exp(\frac{-\gamma_{\textrm{21}}D\Gamma}{\Omega_{\textrm{c}}^{2}+4\gamma_{31}\gamma_{21}}) \\
\times\exp(-2(\ln2)(\frac{t-T_{\textrm{d}}}{\beta T_{\textrm{p}}})^{2}))^{2} dr,
\end{split}
\end{equation}
where the factor

\begin{equation}
\begin{aligned}
\beta=\sqrt{1+\frac{32(\ln2)D\Gamma(\gamma_{31}\Omega_{\textrm{c}}^{2}+2\gamma_{31}^{2}\gamma_{21}+4\gamma_{21}^{3})}{T_{\textrm{p}}^{2}(\Omega_{\textrm{c}}^{2}+4\gamma_{31}\gamma_{21})^{3}}},
\end{aligned}
\end{equation}
and the group delay $T_{\textrm{d}}=D\Gamma(\Omega_{c}^{2}+4\gamma_{21}^{2})/(\Omega_{\textrm{c}}^{2}+4\gamma_{31}\gamma_{21})^{2}$. The normalized Rabi frequency of the probe pulse is assumed to have a Gaussian distribution $\Omega_{\textrm{p}}$=$\exp$⁡(-$r^{2}$/$R^{2}$), where $r$ is the radial coordinate of the fibre core and $R$=22 $\mu$m is the 1/$e^{2}$ mode field radius. The ground state coherence $\gamma_{21}$, which is mainly due to the inhomogeneous differential ac Stark shift of the control and dipole beams, is fixed at 2$\pi\times$(2$\times$10$^{5}$ Hz) based on the measurement of EIT linewidth and ground and excited states coherence $\gamma_{31}$ is fixed at half of the $^{85}$Rb D1 line $P$ state decay rate $\Gamma$= 2$\pi\times$(5.75$\times$10$^{6}$ Hz). The optical depth $D$ then follows the Gaussian distribution $D=L\times n\times \exp⁡(-2r^{2}/R^{2})\times \sigma \times \exp⁡(-2r^{2}/R^{2})$, where the length of the ensemble $L$ along the fiber axis is approximated with 5 mm \cite{Xin}, the number density $n$ is approximated with 1.3$\times$10$^{11}$ cm$^{-3}$ from the maximum measured $D$=40, and $\sigma$=0.65$\times$10$^{-9}$ cm$^{2}$ is the scattering cross-section of the D1 line $F$ = 3, $m$ = 0 to $F$' = 3, $m$ = $\pm$1 transition at the center of the dipole trap.

When the cloud size in the radial direction is assumed to have the same distribution as the light in the fiber, the simulated delayed pulse is within 10$\%$ of the experimental delayed pulse in terms of delay time and efficiency. The uniform profile calculation uses the peak intensity of control and probe beams, and the peak optical depth. The ground state coherence $\gamma_{21}$ is chosen as 2$\pi\times$(10$^{3}$ Hz) from the coherence time measurement of the microwave Ramsey interferometer before light shift cancellation of the optical lattice described in the following section.

Figure 2(b) compares the group delay of the probe pulse in the fiber with the free space scenario of uniform control beam, probe beam, and atoms density profiles. The inhomogeneous profiles of the control and the atom density in the fiber result in more than a factor of 2 reduction of the group delay and the efficiency. Figure 2(c) shows the temporal profile of the probe pulse after turning off the control beam when it is propagating inside the medium and turning on the control beam again after a storage time $T$. The retrieved pulse has a FWHM duration of 160 ns, which corresponds to about 1 MHz bandwidth. We are able to store and retrieve 10$\%$ of the input probe pulse in the fiber.

For the delayed light measurements, we use a moving optical lattice to guide the atoms into the hollow-core fiber and stop them inside the fiber with a stationary optical lattice. This is to avoid loss and decoherence of the atoms due to their motion in the fiber. The moving optical lattice beams are formed by a pair of counter-propagating fields (100 mW each) in the fiber with a velocity $v$=$\Delta f\times\lambda/2$=2.05 cm s$^{-1}$, where $\Delta f$=50 kHz is the frequency detuning of the two lattice fields, and $\lambda$= 821 nm is the lattice wavelength. We transport about 10$\%$ of the atoms into the fiber compared to the loading with just a single optical dipole beam. The dominant loss is due to heating of the atoms during the transporting process which results in escape of atoms  trapped in the higher vibrational energy levels of the moving optical lattice. Figure 3(a) shows the storage efficiency of the light as a function of time. We plot the efficiency of 1550 nm light propagating in a solid-core fiber loop for comparison. Our result exceeds the performance of 1550 nm light after 600 $\mu$s of delay time.

\begin{figure*}
\includegraphics[scale=0.8]{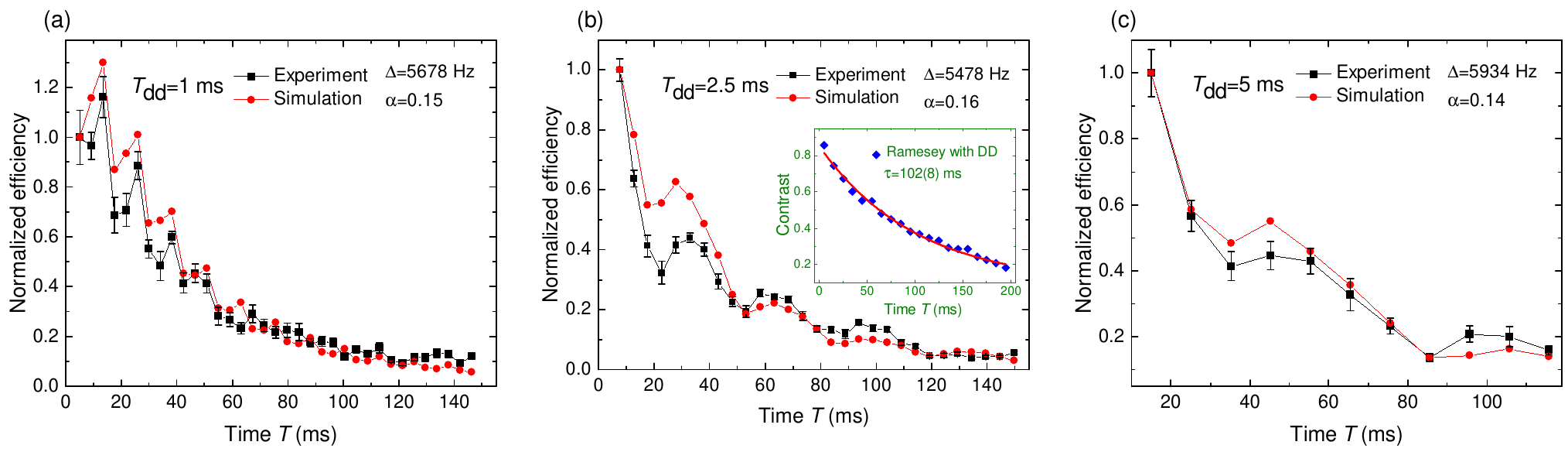}
\caption{Efficiency versus delay time with different dynamic decoupling (DD) pulse separation times. The black squares represent an average of 40 experimental runs of light delay efficiency and the error bars indicate the standard errors. The red circles are the fit of the simulation to the experimental data. The lines are used to guide the eyes. The efficiencies are normalized to the first points for each $T_{\textrm{dd}}$. The ratio of the Gaussian waists of the ensemble to the control beam in radial direction $\alpha$ and the frequency difference of the microwave and the EIT two-photon detuning $\Delta$ are set as free parameters to fit the data. Due to the radial distribution of the EIT beams and atoms, the Rabi frequencies of the control and the probe pulses are set with a Gaussian distribution in the radial direction in the simulation, where their maximum Rabi frequencies excite a $\pi$/2 pulse with 250 ns duration. The duration of the microwave $\pi$ pulses is 37 $\mu$s and the Rabi frequency is assumed uniform across all the atoms. The inset shows the measurement of contrast versus time using microwave Ramsey interferometer with DD sequence. The curve is an exponential fit to the experimental data.}
\end{figure*}

We further improve the performance by minimizing the differential ac Stark shift of $|1\rangle$ and $|2\rangle$ states which is the primary decoherence source of the atomic spin coherence \cite{Xin}. This is carried out by introducing an elliptical-polarized dipole beam and an external magnetic field \cite{Der} to create a vector light shift to cancel the scalar light shift. Figure 3(b) shows the storage efficiency as a function of the delay time. We apply an external magnetic field of 2.08 G and optimize the efficiency by changing the ellipticity of lattice beams through quarter-wave plates. An exponential decay function is fitted to the data with the 1/$e$ decay time of 20 ms, which is limited by the inhomogeneous stray magnetic field of the experimental apparatus \cite{Xin}. Figure 3(c) shows experimental results of the normalized efficiency of light storage in a free-falling atomic spin-wave for comparison. The storage time is shorter than for the atoms trapped in a stationary optical lattice. We attribute this to the inhomogeneity of the magnetic field and dipole potential along the trajectory of atoms that perturb the spin coherence and the relative motion of atoms, respectively.

One way to prolong the delay time is to execute the dynamic decoupling (DD) method \cite{Sagi,Dud}, a series of $N$ microwave population inverting $\pi$ pulses separated by cycling time $T_{\textrm{dd}}$ is applied to the atoms during the storage period to rapidly rephase the spin coherence, where $N$ is an even integer. We study the DD method under our fiber delay line condition experimentally and theoretically. Figure 4 shows the delay efficiency versus time measurements with $T_{\textrm{dd}}$=1, 2.5, and 5 ms. We observe the oscillation of the efficiency over time due to the frequency difference $\Delta$ between the microwave frequency and the two-photon detuning between the control and the probe pulses. We confirm the oscillation by simulating the rotation of the Bloch vectors of the atoms under DD.

In the simulation, we set the residual inhomogeneous broadening of the dipole beam $\delta$=2$\pi\times$(50 Hz) from the data obtained in Fig. 3(b) and introduce $\alpha$ as the ratio of Gaussian waists of the atomic ensemble to the control beam in radial direction and set it as a free parameter together with $\Delta$. We fit the simulation to the data and find the parameters $\alpha$ and $\Delta$ for each $T_{\textrm{dd}}$. The decay of the efficiency is due to the spatial inhomogeneity of the control beam propagating in the fiber, which is determined by $\alpha$. Further details can be found in the Appendix.

In the Bloch sphere picture, the inhomogeneous control pulse prepares Bloch vectors on the Bloch sphere with a distribution for the DD sequence. To verify it, we use the same DD sequence on a microwave Ramsey interferometer, replacing the EIT probe and control pulses with microwave pulses which are assumed to be homogeneous across the whole atomic ensemble. We find that the 1/$e$ decay time $\tau$ of the contrast of the pure microwave sequence is two times longer than the storage time under the same DD; see the inset of Fig. 4(b).

Light storage in the hollow-core fiber shows three orders of magnitude slower decay rate than the propagation loss of light in the solid-core fiber. Moreover, the delay time could be varied precisely by merely turning on and off the control pulse as an optical switch without modifying any hardware as in the solid-core fiber loop delay line. The low initial storage efficiency ($\sim3\%$) in our experiment compared to the free space case is due to the inhomogeneous profile of the control beam and atom density inside the fiber. We have achieved the time-bandwidth product (TBP) to about 5$\times$10$^{4}$, a figure of merit for quantifying the performance of the optical storage capacity. The solid-core fiber systems \cite{Zhu,Sag,Jin,Sag2} have achieved TBP to about 800 with a few percent of efficiency and atomic systems \cite{Spr,Gou,Say,Cor,Bla,Pet} have achieved TBP to about 10 with about 30$\%$ efficiency. Our efficiency can be improved by increasing the optical depth \cite{Bla,Hsi} or using other light storage methods such as Autler$-$Townes splitting to ease the demand of high optical depth \cite{Sag3}. The near-infrared wavelength of our light can be converted into the telecom band in the same medium using a higher level of atomic transitions \cite{Cha,Rad} and be extended to single photons for quantum network applications \cite{Kim}.

We thank Alex Kuzmich and Yi-Hsin Chen for reading the manuscript. This work is supported by the Singapore National Research Foundation under Grant No. NRFF2013-12, Nanyang Technological University under start-up grants, and Singapore Ministry of Education under Grants No. Tier 1 RG107/17.

\appendix*
\section{Details of simulations of the light storage efficiency under dynamical decoupling}
    \subsection{\label{assumptions}Assumptions}
We assume the following for the dynamical decoupling (DD) simulations:
\begin{enumerate}
\item{The effective Rabi frequency $\Omega_R(r)$ of the EIT control and probe pulse for an atom at radial position $r$ is Gaussian distributed according the fundamental mode field profile of the hollow-core fiber as $\Omega_R(r) = \Omega^{max}_R e^{-2r^2/R^2}$, where $R = 22~\mu$m is the $1/e^2$ mode field radius and $\Omega^{max}_R = \pi/(500~\mathrm{ns})$.}

\item{The EIT fields are two-photon resonant on the unperturbed atomic transition.}

\item{The Rabi frequency $\Omega_M$ of the microwaves is uniform across the ensemble with a $\pi$-pulse time of 37~$\mu$s.}

\item{The microwave detuning from the unperturbed transition frequency is $\Delta$. $\Delta$ is used as one of the fit parameters to the data.}

\item{The inhomogeneous broadening $\delta(r)$ caused by the residual optical lattice differential ac Stark shift is also Gaussian distributed as $\delta(r) = \delta_{max} e^{-2r^2/R^2}$, where $\delta_{max} = 2\pi \times (50~\mathrm{Hz})$ from the data in Fig. 3(b) of the main paper.}

\item{Initially, at $t=0$, the phase of the microwaves is the same as the EIT fields, i.e., over time $t$, the microwave phase relative to an atom at position $r$ is $(\Delta+\delta(r))t$.}

\item{The atoms are Gaussian distributed according to $\rho(r) = \mathrm{exp}\left( -r^2/2R^2_{atom}\right)/R_{atom}^2$ where $R_{atom}$ is the rms radius of the atomic ensemble. The probability distribution is normalized:  $\int_0^\infty \rho(r) r dr = 1$. In the simulation, the atom positions are fixed for the duration of the dynamical decoupling sequence. The ratio $\alpha$ of the ensemble size to the mode field radius, i.e., $\alpha \equiv R_{atom}/R$, is used as the other fit parameter to the data.}
\end{enumerate}

\subsection{\label{single atom}Simulation on a single atom}
The goal is to determine the collective state of the atoms after the EIT pulse and microwave dynamical decoupling pulses with the atoms evolving phase in the optical lattice.

Beginning with an atom at position $r$ in the ground state, represented by the Bloch vector $\mathbf{u}_0 = (0,0,-1)$, we model the EIT pulse as a two-photon Raman $\pi/2$-pulse acting on the Bloch vector. The rotation matrix representing the $\pi/2$ pulse takes into account the inhomogeneous Rabi frequency $\Omega_R(r)$ and detuning $\delta(r)$. In the Bloch sphere picture, the inhomogeneous EIT pulse prepares Bloch vectors on the Bloch sphere with a distribution for the DD sequence.

After the EIT pulse, the atom evolves phase $\delta(r) T_{\mathrm{dd}}$ in the optical lattice for DD cycling time $T_\mathrm{dd}$. In the Bloch sphere picture, the Bloch vector rotates about the $z$-axis by angle $\delta(r) T_{\mathrm{dd}}$. A microwave $\pi$-pulse is then applied, with Rabi frequency $\Omega_M$, detuning $\Delta + \delta(r)$, and phase $(\Delta + \delta(r))t$, where $t$ is the time at which the $\pi$-pulse is applied. Again, the radial dependence encodes the inhomogeneity of the ensemble with respect to the microwaves. The DD sequence repeats the elementary sequence [free-evolution $T_\mathrm{dd}$ -- $\pi$-pulse] for an even number of times. The Bloch vector of the atom $\mathbf{J} = (J_x, J_y, J_z)$ is numerically determined after each repeating unit of the DD sequence.



\subsection{\label{weighting}Weighting over ensemble}
The ensemble has a radial distribution across the fiber mode profile leading to a distribution of Rabi frequency $\Omega_R(r)$ and detuning $\delta(r)$. As there is a one-to-one mapping between the radial position $r$ and the normalized fiber mode intensity $I(r) = e^{-2r^2/R^2}$, the single atom simulation $\mathbf{J}$ is actually a function $\mathbf{J}(I)$ of the normalized mode intensity $I$.

To simulate the effect of this inhomogeneous distribution, we weight the simulated single atom Bloch vectors $\mathbf{J}(I)$ by the intensity distribution $\rho(I)$ that the ensemble samples as in the following
\begin{equation}
\langle \mathbf{J}\rangle  \equiv \int^1_0 \mathbf{J}(I) \rho(I) dI	.	\label{eq1}
\end{equation}
This defines the collective Bloch vector $\langle \mathbf{J}\rangle $ that we use to calculate the retrieval efficiency. Using the Gaussian atom distribution provided in the assumptions, it can be shown that the equivalent intensity distribution is given by
\begin{equation}
\rho(I) = \frac{1}{4\alpha^2}   I^{\frac{1}{4\alpha^2} -1}  .		\label{eq2}
\end{equation}
For the numerical simulations, we approximate the integral using the trapezium rule.

\subsection{\label{efficiency}Retrieval efficiency}
The retrieval efficiency is determined by the collective coherence of the ensemble, which corresponds to the component of the collective Bloch vector in the equatorial plane $J_\perp^2 \equiv \langle J_x \rangle^2 + \langle J_y \rangle^2$, where $\langle J_{x} \rangle$ is the $x$-component of the collective Bloch vector $\langle \mathbf{J} \rangle$; similarly for $\langle J_y \rangle$. The normalized efficiency is then $(J_\perp/J_{\perp, 0})^2$ where $J_{\perp,0}$ is the collective coherence of the ensemble right after the EIT storage process.


\subsection{\label{fitting}Fitting simulation to data}
To fit the simulation to the data, we use the method of least squares. Specifically, we define the sum of squares deviation between the simulation and experimental data points as
\begin{equation}
\mathrm{SS} = \sum_i \frac{  (\mathrm{simulation(\Delta,\alpha)}_i - \mathrm{data}_i)^2  }{  \mathrm{error}_i^2  }	\label{eq2}
\end{equation}
where error$_i$ is the standard error of the $i$-th experimental data point. We then numerically minimize $\mathrm{SS}$ over the free parameters $\Delta$ and $\alpha$ to obtain fits to the data sets.


\nocite{*}

\bibliography{apssamp}

\end{document}